\documentclass[preprint,showpacs,preprintnumbers,amsmath,amssymb,floatfix,pre]{revtex4}

\usepackage{graphicx}
\usepackage{dcolumn}
\usepackage{bm}
\usepackage{verbatim}
\def\hf{{\frac{1}{2}}}
\def\d{\mathrm{d}}

\begin{document}

\title{Runaway expansion in confined quasi-2D plasmas and vortex fluids}
\author{T. D. Andersen}
\email{andert@alum.rpi.edu}
\noaffiliation
\date{\today}


\begin{abstract}
The confined, quasi-two-dimensional guiding center plasma and a system of interacting line vortices in an ideal fluid are examples of Hamiltonian systems with infinite interaction distances.  The existence of metastable states with negative specific is investigated by standard entropy maximization of the thermodynamic limit of vortices as they become infinitesimal and form a continuous field.  We find metastable states and suggest that these imply a runaway reaction leading to a rapid expansion of a confined plasma or fluid similar to the rapid collapse of globular clusters in astrophysics.
\end{abstract}

\maketitle


\section{Introduction}
\begin{figure}
\begin{center}
\includegraphics[width = 0.6\textwidth]{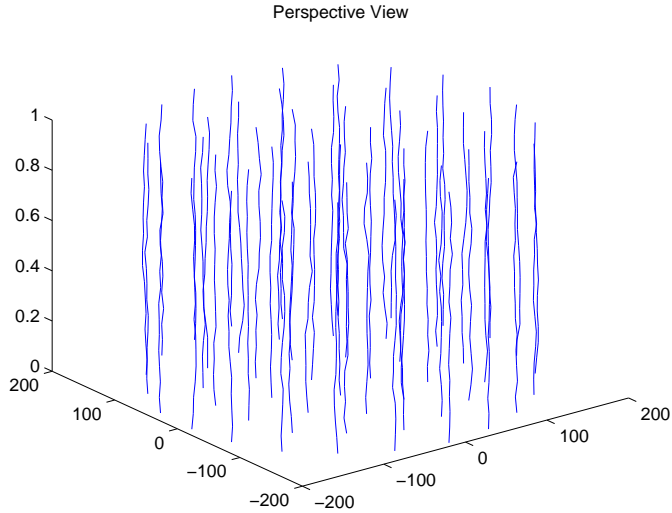}
\end{center}
\caption{This output from a Monte Carlo simulation of the energy functional.}
\label{fig:paths3d}
\end{figure}

There has been considerable interest recently in metastable states and negative specific heat, particularly related to fluids and plasmas \cite{Kiessling:2003, Andersen:2007b}.  The guiding center plasma or ideal fluid vorticity model for quasi-2D columns of electrons or lines of vorticity is similar to the widely studied two-dimensional model but the lines contain small variations which can change the dynamics of the system.  An ensemble of filaments $\{\vec{\phi}_1(\tau),\dots,\vec{\phi}_N(\tau)\}$,
\begin{equation}
E_N[\vec{\phi}_1\dots\vec{\phi}_N] = \sum_i \int_0^{l} \d\tau \frac{\alpha\Gamma}{2}\left|\frac{d\vec{\phi}_i}{d\tau}\right|^2 - \frac{1}{\epsilon}\sum_{i<j} \Gamma^2\log|\vec{\phi}_i - \vec{\phi}_j|,
\end{equation} where $1/\epsilon$ is the coupling constant, all the filaments have the same average circulation, $\Gamma$, $\alpha$ is a core elasticity constant related to the frequency, and $l$ is the length of the period under periodic boundary conditions $\vec{\phi}_i(0)=\vec{\phi}_i(l)$.  Each filament $\vec{\phi}_i(\tau)=(x_i(\tau),y_i(\tau))$ is a vector in the plane with a parameter $\tau$ representing the third dimension.  This is under special asymptotic assumptions that the filaments are nearly parallel and far enough apart \citep{Lions:2000}.  Because of the nearly parallel assumption, this model neglects vortex stretching which is assumed to be too small to affect the statistics.   An example of such an ensemble is pictured in Figure \ref{fig:paths3d}.  

There is a significant difference between the two-dimensional one-component Coulomb plasma and this quasi-2D model.  Because the total energy is entirely dependent on how far apart the lines are, an ensemble of two-dimensional lines at a fixed energy has a maximum radius beyond which the lines cannot move while the quasi-2D lines can move, theoretically, as far apart as there is space available because the potential energy can always be balanced by altering the kinetic energy.  In addition, we show in this paper that metastable states exist in the quasi-2D model that do not exist in the 2D model.

Statistical derivations for three dimensional fluids of any kind have tended to focus on single vorticity columns \cite{Hasimoto:1972,
Callegari:1978} and \cite{Klein:1991,Ting:1991} or statistical treatments of ensembles of two-dimensional point vortices \cite{Onsager:1949, Joyce:1973, Edwards:1974, Chorin:1994}.  The nearly parallel vortex filament model of \cite{Klein:1995, Lions:2000} for Navier-Stokes fluids is an exception. 

The nearly parallel model can be extended to electron columns by a generalized vorticity model for electron plasmas \citep{Gordeev:1994, Uby:1995, Kinney:1993} which takes the magnetic and electric fields into account as well as the vorticity.  In the case of charged particles, the vorticity must, essentially, be gauge invariant: the magnetic field, $-e\vec{B}/m = (-e/m)\nabla\times \vec{A}$, and the charged fluid vorticity, $\vec{\omega} = \nabla\times\vec{v}$,  combine into a general vorticity field $\Omega = m^{-1}\nabla\times \vec{p}$ where the generalised or ``canonical'' gauge invariant momentum is, $\vec{p} = m\vec{v} - e\vec{A}$, $m$ is the electron mass, $-e$ is the electron charge, $\vec{v}$ is the fluid velocity field, and $\vec{A}$ is the magnetic vector potential field.  The generalized angular momentum is $\vec{L}=\vec{r}\times \vec{p}$.

Electron column core sizes are small, equal to the Larmor radius, $a$, about a tenth of a millimeter for electrons, hence the core structure may be abstracted by the local induction approximation (LIA). The size of the local region, which is the wavelength of the highest energy frequencies on the filaments, is naturally given by the London wavelength, $b=c/\omega_{pe}$, where the electron plasma frequency is given by $\omega^2_{pe} = ne^2/m\epsilon_0$ and $n$ is the electron number density of the individual filament \cite{Uby:1995}.  Because the filaments are all nearly parallel, collisions between points not in the same cross-section are small, and the two dimensional Coulomb interaction gives the potential to leading order.

\section{Method}
A filament nearly parallel to the $z$ axis has a $C^2$ curve $\vec{\phi}(\tau)=(x(\tau),y(\tau))$ with $L^2_{[0,l]}$ derivative where $l\ll 1$ is a small length scale.  Assuming no short-wave disturbances (knots, kinks, etc.), the filament has a LIA kinetic energy functional,
\begin{equation}
E_1[\vec{\phi}] = \int_0^l \d\tau \frac{\alpha\Gamma}{2}\left|\frac{d\vec{\phi}}{d\tau}\right|^2
\end{equation}  where $\Gamma$ is the generalized circulation of the filament \cite{Uby:1995} and $\alpha=\log(b/a) + 1$, $b$ is the arclength of the ``local'' region where the induction takes place, i.e. the wavelength of the highest frequency modes on the filament, and $a$ is the core size of the filament.  The ratio $b/a$ does not vary significantly over the length of the filament; therefore, it can be assumed to be constant \cite{Ting:1991}.  The core size $a$ of electron filaments is the Larmor radius while a natural choice for $b$ is the London wavelength, $b=c/\omega_{pe}$, where the electron plasma frequency is given by $\omega^2_{pe} = ne^2/m\epsilon_0$ and $n$ is the electron number density of the individual filament \cite{Uby:1995}.

For $N$ such filaments, they have, under suitable asymptotic assumptions such that the filament core size is much smaller than the intervortex spacing, a 2-D Coulomb interaction such that,
\begin{equation}
E_N[\vec{\phi}_1\dots\vec{\phi}_N] = \sum_i \int_0^{l_i} \d\tau \frac{\alpha\Gamma}{2}\left|\frac{d\vec{\phi}_i}{d\tau}\right|^2 - \frac{1}{\epsilon}\sum_{i<j} \Gamma^2\log|\vec{\phi}_i - \vec{\phi}_j|,
\end{equation} where $1/\epsilon$ is the coupling constant and all the filaments have the same average circulation \cite{Kinney:1993}.  The kinetic energy is the square amplitude of the filament with zero amplitude implying zero kinetic energy; hence the filament's self energy is not directly connected to the microscopic temperature.  

The kinetic energy of generalized angular momentum for a single filament of electrons is $A_1=\hf I\bar{\omega}^2$ where $I$ is the moment of inertia of the filament, $I=\int_0^l\d\tau\,|\vec{\phi}|^2$, and $\bar{\omega}$ is the generalized frequency of rotation.  For $N$ filaments, all with the same frequency,
\begin{equation}
A_N[\vec{\phi}_1\dots\vec{\phi}_N] = \hf \mu'\sum_i \int_0^{l_i} \d\tau\, \Gamma|\vec{\phi}_i|^2,
\end{equation} where $\mu'\Gamma=\bar{\omega}^2$.  Now choose mass and charge units such that $e=m=1$ for electrons.

No boundary conditions are imposed perpendicular to $z$ since the plasma is fully contained by the magnetic field and never contacts any surfaces.  The confinement radius $R$ is a bounding radius representing the distance from the center of the plasma to the outer edge where the density falls to zero (sometimes abruptly).  This is smaller than the radius of the container.  Provided the magnetic surfaces to which they are confined are closed, the filaments cannot interact with material surfaces.

In the thermodynamic limit as $N\rightarrow\infty$ such that total circulation, $\Lambda = \Gamma N$, is constant, the filaments have area density $f(\vec{c},\vec{r},\tau)=\Gamma g(\vec{c},\vec{r},\tau)$ (where $\vec{c}=d\vec{\psi}/d\tau$ and $\Gamma$ is the generalized circulation) such that, if $\sigma=(\vec{c},\vec{r})$, the kinetic energy is,
\begin{equation}
\mathcal{T} = \hf\int \d^4\sigma \d\tau\, f\alpha c^2 ,
\label{eqn:ken}
\end{equation} where $c=\|\vec{c}\|$.  In a rotating frame the kinetic energy of angular momentum becomes a potential added to the interaction potential,
\begin{equation}
\mathcal{V} = \hf\mu' \int \d^4\sigma \d\tau f |\vec{r}|^2 -\frac{1}{\epsilon} \int \d^4\sigma \d^4\sigma' \d\tau\, ff'\log|\vec{r} - \vec{r'}|,
\label{eqn:pen}
\end{equation} where $f' = f(\vec{c'},\vec{r'},\tau)$ and particle number is $N=\int \d^4\sigma \d\tau\, g$.  All integrals are over the interior of the torus.  Since the energy functional does not depend on $\tau$, the density with maximal entropy does not depend on $\tau$ either (as one can show from the variation); therefore, we drop the integrals over $\tau$ and assume the area density is constant in $\tau$.  All functionals are now per unit length.

For a fixed energy system, the entropy (with Boltzmann's constant $k_B=1$),
\begin{equation}
S = - \int \d^4\sigma\, g\log g,
\label{eqn:entropy}
\end{equation} is maximal in the most-probable macrostate.

To maximize the entropy, we must solve the variational problem,
\begin{equation}
\delta S = 0,
\end{equation} subject to the constraints, $\mathcal{T}+\mathcal{V}=E$, $N$ fixed.  That is for some small parameter $\lambda$, we define a family of density functions $g(\sigma;\lambda)$, such that entropy is maximal ($\delta S = dS/d\lambda = 0$) at $g(\sigma;0)$.

The method of Lagrange multipliers provides the equation for the variation of $S$ subject to the constraints,
\begin{equation}
\delta S + \beta'\delta E + \nu \delta N = 0,
\end{equation} where $\beta'$ is inverse temperature, $\nu$ is a normalization parameter.  (Angular momentum is automatically conserved by the rotational invariance of the energy.)  Taking the variation (\cite{Lynden:1968}) we have,
\begin{equation}
\log g + 1 + \beta'\Gamma\left(\frac{\alpha}{2}c^2 + \frac{\mu'}{2}r^2 + \psi \right) + \nu = 0,
\label{eqn:var1}
\end{equation} where 
\begin{equation}
\psi(\vec{r}) = -\frac{1}{\epsilon}\int \d\vec{r}' \rho(\vec{r}')\log|\vec{r} - \vec{r}'|
\end{equation} is the 2D Coulomb potential.  Let $H = \frac{\alpha}{2}c^2 + \frac{\mu'}{2}r^2 + \psi$.

Solving \ref{eqn:var1} for the density,
\begin{equation}
g = Ae^{-\beta'\Gamma H},
\label{eqn:totaldens}
\end{equation} where $A = \exp[-(\nu+1)]$ is a normalization constant that gives the particle number $N$.  Let $\beta=\beta'\Gamma$.  Showing the equipartition of energy, the average kinetic energy per filament is
\begin{equation}
\frac{\int \d^2 c\, f \frac{\alpha}{2} c^2}{\int \d^3c\, f} = \frac{\int \d^2c\, \exp\left(-\hf\beta\alpha c^2\right)\hf \alpha c^2}{\int \d^2 c \exp\left(-\hf \beta\alpha c^2\right)} = \frac{1}{\beta}.
\end{equation}

To solve for the spatial density, $\rho$, we can integrate equation \ref{eqn:totaldens} over all ``velocities'', $\vec{c}$,
\begin{equation}
\rho = \int \d^2c\,f = B\exp[-\beta(\psi + \mu' r^2/2)],
\end{equation}  where $B = A(2\pi/\beta)$.

Replacing $\rho$ in the potential with the above equation gives an integral equation for the most-probable potential inside the circle,
\begin{equation}
\psi(\vec{r}) = -\int \d\vec{r'} Be^{-\beta(\psi(\vec{r}') + \mu' r^2/2)}\log|\vec{r} - \vec{r}'|.
\end{equation}  This integral equation is equivalent to the Poisson equation,
\begin{equation}
\nabla^2\psi(\vec{r}) = -\frac{4\pi}{\epsilon}\rho = \left\{\begin{array}{lr} -4\pi B\exp[-\beta(\psi + \frac{\mu'}{2}r^2)]/\epsilon & |\vec{r}|<R \\ 0 & |\vec{r}|\geq R \end{array}\right.
\label{eqn:pot}
\end{equation} with boundary conditions such that $\psi$ and $d\psi/dr$ are continuous at the boundary $r=R$.  Because of the axisymmetry of the energy the potential must also be statistically axisymmetric.  Converting to polar coordinates and \ref{eqn:pot} gives the ODE,
\begin{equation}
\frac{1}{r}\frac{d}{dr}\left(r\frac{d\psi}{dr}\right) = -\frac{4\pi}{\epsilon} Be^{-\beta(\psi(r) + \mu' r^2/2)},
\label{eqn:ode1}
\end{equation} for $r<R$.

Because the potential is repulsive, the only solutions to this equation have finite density everywhere (unlike gravitational systems, infinite densities such as black holes are not possible in a repulsive Coulomb system).  To simplify, we make the change of variables, $v_1 = \beta(\psi - \psi(0))$ and $r_1 = \sqrt{4\pi\beta A\exp[-\beta\psi(0)]/\epsilon}\cdot r = \sqrt{4\pi\beta\rho(0)/\epsilon}\cdot r$, which simplifies the ODE (\ref{eqn:ode1}) to,
\begin{equation}
\frac{d^2v_1}{dr_1^2} + \frac{1}{r_1}\frac{dv_1}{dr_1} + e^{-v_1 - \beta\mu' r_1^2 (R^2/z^2)/2} = 0,
\label{eqn:ode}
\end{equation} where $z=\sqrt{4\pi\rho(0)\beta/\epsilon}R$ and with boundary conditions,
\begin{equation}
v_1(0) = v_1'(0) = 0,
\end{equation}  where $v_1'=dv_1/dr_1$.  Note that the plasma density can be written $\rho = \rho(0)e^{-v_1}$ or ($v_1 = -\log[\rho/\rho(0)]$); therefore, the variable $v_1$ describes how the density changes as the distance from the origin changes and decreases monotonically from 0 at $r_1=0$.   

The viral theorem of Clausius may be applied,
\begin{equation}
2\mathcal{T} + \mathcal{V} = 3pV,
\end{equation} where
\begin{equation}
p = \int_{r=R} \d^2c f \frac{1}{3} \alpha c^2,
\end{equation} is the surface ``pressure'' and $V=\pi R^2$ is the area of the circle.  From the equipartition theorem, where $\Lambda$ is the total circulation,
\begin{equation}
\mathcal{T} = \frac{\Lambda}{\beta}.
\label{eqn:kinen}
\end{equation} Using the Virial theorem,
\begin{equation}
E = 3pV - \mathcal{T} = 3pV - \frac{\Lambda}{\beta},
\label{eqn:en}
\end{equation} and
\begin{equation}
\mathcal{V} = 3pV - 2\frac{\Lambda}{\beta}.
\end{equation}

Evaluating the surface pressure,
\begin{equation}
p = \int_{r=R} \d^2c\,f \frac{1}{3} \alpha c^2 = \frac{2}{3}\frac{1}{\beta}\int_{r=R} \d^2c\, f = \frac{2\rho(R)}{3\beta},
\label{eqn:pressure}
\end{equation} where 
\begin{equation}
f = A\exp\left[-\beta\left(\alpha\frac{c^2}{2} + \mu' \frac{r^2}{2} + \psi\right)\right].
\end{equation}  

Integrating \ref{eqn:ode1}, circulation $\Lambda$ is given by,
\begin{equation}
\frac{\Lambda}{\epsilon} = -\left(r\frac{d\psi}{dr}\right)_{r=R} = -\frac{1}{\beta}\left(r_1\frac{dv_1}{dr_1}\right)_{r_1=z}.
\end{equation}  Then
\begin{equation}
\beta = -\frac{\epsilon zv_1'(z)}{\Lambda}
\label{eqn:beta}.
\end{equation} For the rest of this section, $v_1'$ and $v_1$ shall refer to $v_1'(z)$ and $v_1(z)$ only.

The pressure is,
\begin{equation}
p = \frac{2}{3}\frac{\rho(R)}{\beta} = \frac{2}{3}\frac{\rho(0) e^{-v_1 - \beta\mu R^2/2}}{\beta}.
\end{equation}  Since $r_1^2/r^2 = |4\pi\beta\rho(0)/\epsilon|$, and eliminating instances of $\beta$ with \ref{eqn:beta},
\begin{equation}
p = \frac{2}{3}\frac{\epsilon z^2}{4\pi R^2}\frac{e^{-v_1 - \beta\mu' R^2/2}}{\beta^2} = \frac{\Lambda^2}{6\pi R^2\epsilon}\frac{z^2e^{-v_1 + \mu' zv_1'\epsilon R^2/(2\Lambda)}}{(-zv_1')^2}.
\end{equation}   Let $\mu=\epsilon \mu' R^2/(2\Lambda)$, then
\begin{equation}
3pV = \frac{\Lambda^2}{2\epsilon}\frac{z^2e^{-v_1 + \mu zv_1'}}{(-zv_1')^2},
\end{equation} and the energy from \ref{eqn:en} is
\begin{equation}
E = \frac{\Lambda^2}{\epsilon}\left(\frac{z^2e^{-v_1 + \mu zv_1'}}{2(-zv_1')^2} - \frac{1}{(-zv_1')}\right).
\label{eqn:finalen}
\end{equation}

The entropy, obtained from \ref{eqn:entropy}, is given by,
\begin{equation}
S = \frac{1}{\Gamma}\left\{\beta \left(E - \frac{\Lambda^2}{\epsilon}\log R\right) - \Lambda \log p\beta^2 + \Lambda \log 2\pi\Gamma\right\}.
\end{equation}

The specific heat is,
\begin{equation}
c_v = \frac{dE}{dT} = \frac{\frac{dE}{dz}}{\frac{dT}{dz}}.
\label{eqn:specheat}
\end{equation}

When $\mu zv_1'r^2/R^2=v_1(z)$ we have a constant density solution, where the potential energy and the confinement are perfectly balanced.  For $\mu zv_1'r^2/R^2>v_1(z)$ the potential dominates, and the density tends to favor expansion with higher density toward the wall of the container.  These profiles, although stable, are not suitable for containment.  For $\mu zv_1'r^2/R^2<v_1(z)$ the magnetic confinement dominates and a more Gaussian density profile is preferred with higher density toward the middle.  These profiles are useful for containment.

Numerically evaluating $v_1$ and $v_1'$ for a range of $z$ values (Fig. \ref{fig:negspec}) in the strong rotation regime $\mu>0.5$, we have negative specific heat.  At $\mu=0.5$ the energy is zero indicating that the potential energy, which is the only energy component that can take negative values, perfectly balances the kinetic energy and the confinement potential; at this point the specific heat is zero.  In the weak rotation regime with $\mu<0.5$, the specific heat remains positive (Fig. \ref{fig:negspec2}).

\section{Results}
\begin{figure}[htb]
\centering
\includegraphics[width=0.6\linewidth]{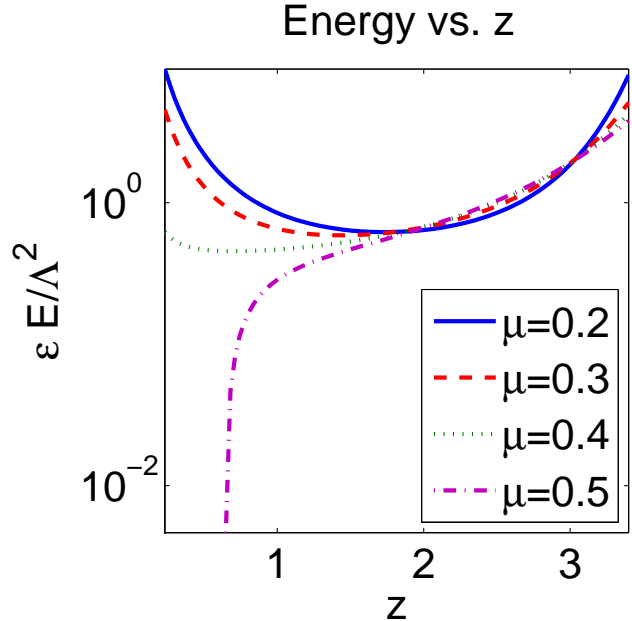}
\label{fig:negspec}
\caption{For regimes with $\mu<0.5$, confinement is weak enough that stable states exist, indicated by energy decreasing with $z$, but beyond that all states are metastable, indicating that equilibrium is not possible.  When energy increases with $z$, the system is metastable, and, because this leads the system to run away to hotter temperatures where the central density decreases, runaway expansion results.}%

\end{figure}

\begin{figure}[htb]
\centering
\includegraphics[width=0.6\linewidth]{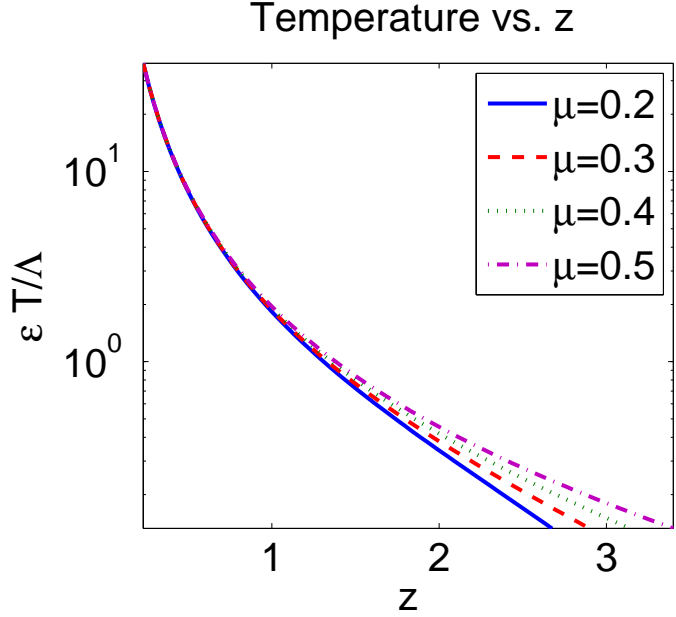}
\caption{Temperature decreases with $z$.  Therefore, if energy increases with $z$, the specific heat is negative.  If energy decreases with $z$ the system has stable equilibria.}%
\label{fig:negspec2}
\end{figure}

\begin{figure}[htb]
\centering
\includegraphics[width=0.6\linewidth]{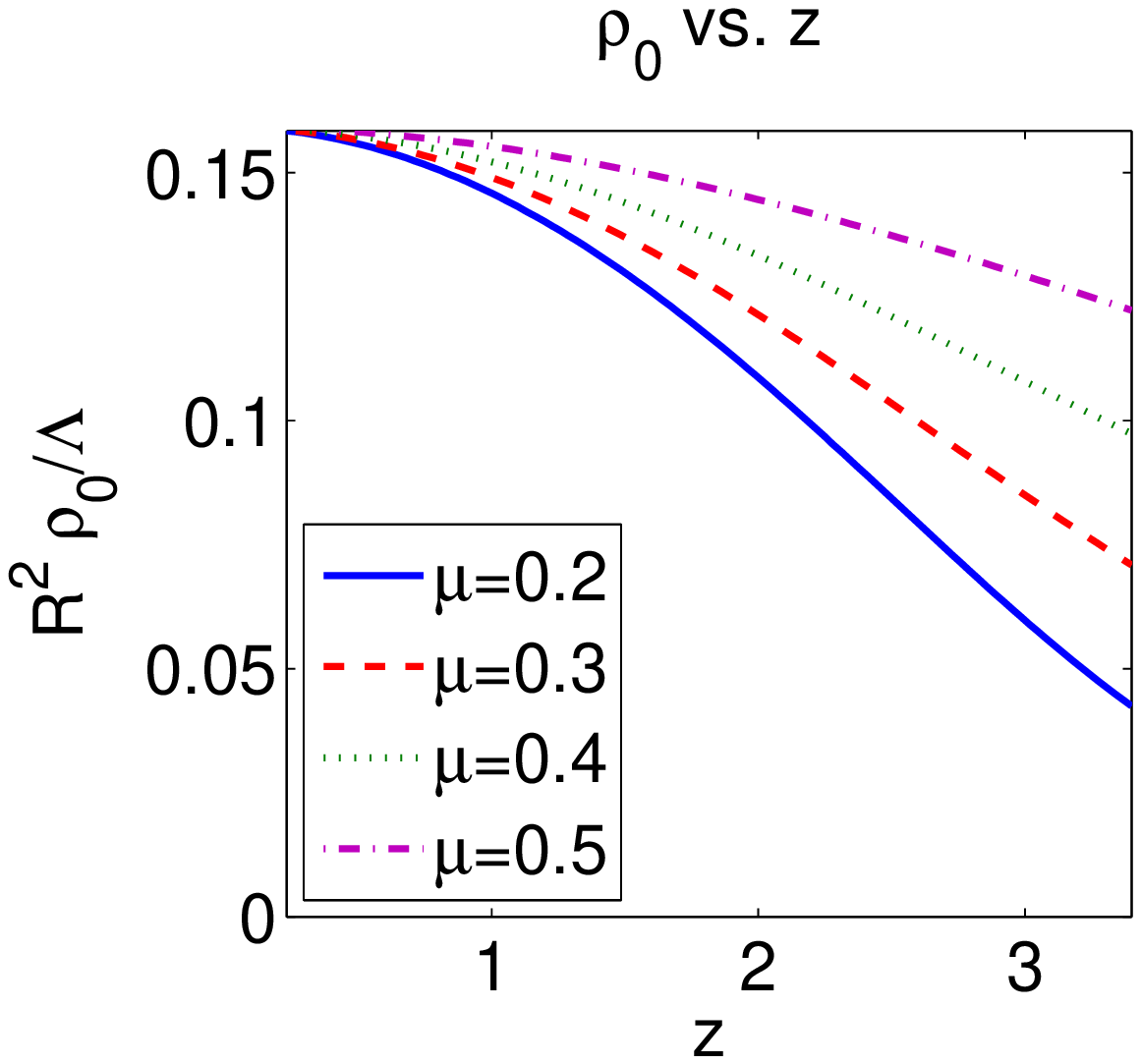}
\caption{As $\mu$ increases, confinement becomes stronger and the central density decreases less with increasing $z$, but, because it is decreasing in the metastable regime, this indicates a core expansion.}%
\label{fig:negspec3}
\end{figure}

Entropy maximization of the model yields the following results:
\begin{enumerate}
\item An expression for the energy in terms of $z$, $v_1(z)$, and $v_1'(z)$,
\begin{equation}
E = \frac{\Lambda^2}{\epsilon}\left(\frac{z^2e^{-v_1 + \mu zv_1'}}{2(-zv_1')^2} - \frac{1}{(-zv_1')}\right),
\label{eqn:finalenres}
\end{equation} where $\Lambda$ is the total circulation, $\epsilon$ is the coupling constant for vorticity, and $\mu$ is a parameter determining strength of the angular kinetic energy;
\item An expression of the inverse temperature of the filaments (Boltzmann's constant, $k_B=1$),
\begin{equation}
\frac{1}{T}=\beta = -\frac{\epsilon zv_1'(z)}{\Lambda}
\label{eqn:betares};
\end{equation}
\item An expression for the central density,
\begin{equation}
\rho(0) = \frac{\epsilon z^2}{4\pi\beta R^2},
\label{eqn:rho0res}
\end{equation} where $R$ is the radius of the confinement area;
\item A second ODE governing the density at a particular energy/temperature,
\begin{equation}
\frac{d^2v_1}{dr_1^2} + \frac{1}{r_1}\frac{dv_1}{dr_1} + e^{-v_1(r_1) + \mu r_1^2 v_1'(z)/z} = 0;
\label{eqn:oderes2}
\end{equation}
with boundary conditions,
\begin{equation}
v_1(0) = v_1'(0) = 0,
\end{equation}  where $v_1'=dv_1/dr_1$;
\item And an expression for the density:
\begin{equation}
\rho(r) = \rho(0)\exp(-v_1(r) + \mu r^2 zv_1'(z)/R^2 ),
\label{eqn:density}
\end{equation} where $r = r_1/\sqrt{4\pi\beta\rho(0)/\epsilon}$ is the distance from the center.
\end{enumerate}
From numerical evaluation  of \ref{eqn:oderes2} the energy (\ref{eqn:finalenres}), inverse temperature (\ref{eqn:betares}), and core density (\ref{eqn:rho0res}) are computed as functions of $z$.  These are plotted for several values of $\mu$ (Fig. \ref{fig:negspec}, Fig. \ref{fig:negspec2}, Fig. \ref{fig:negspec3}).

\section{Discussion}
Negative specific heat indicates a metastable state where the entropy, rather than being globally maximal, is at a saddle point or local maximum, and the system evolves out of this state by increasing temperature, where the term ``temperature'' refers to the amount of kinetic energy per unit circulation and not the electron temperature.  Because the specific heat is negative, increasing temperature lowers the total energy.  In gravitational systems where this phenomenon was first described, this results in gravothermal catastrophe where a globular cluster, for example, experiences a core collapse.  The collapse of the core results in reduced gravitational potential and increased stellar velocities.  Thus, the overall energy decreases while the temperature increases.  It has been argued by analogy that negative specific heat in magnetically confined, neutral electron-positron plasmas would result in core collapse leading to the possibility of nuclear fusion as columns of electrons and ions collapse into one another \cite{Kiessling:2003}.  In a previous paper, however, we showed that self-energy causes an anomalous expansion in the mean radius of the density profile and developed a mean-field formula for it, confirming it with Monte Carlo simulations \cite{Andersen:2008b}.  The same approach in a microcanonical ensemble shows that the system exhibits negative specific heat \cite{Andersen:2007b}.   Therefore, while magnetically confined quasi-2D plasmas exhibit negative specific heat indicating metastable energy states where no equilibrium exists, because the potential is repulsive, the instability is a runaway expansion.    

This expansion arises because increases in kinetic energy and decreases in potential energy both cause expansion.  Analyzing Figs. \ref{fig:negspec}, \ref{fig:negspec2}, and \ref{fig:negspec3}, the mechanism works as follows: (1) The potential energy decreases.  If the magnetic confinement is sufficiently weak at the given density, expanding it decreases the potential energy.  (2) The kinetic energy increases in response, but, because the system has expanded, energy is lost at the outer edges of the plasma.  Thus, the filament temperature has increased but energy decreased.  (3) The increased kinetic energy causes the density profile to spread (by increasing the variance), causing further expansion, and returning the cycle to step (1).  As the potential energy continues to decrease, the kinetic energy continues to increase, and the system experiences a runaway expansion directly analogous to the runaway collapse of a gravitational system.  

Unlike in the gravitational systems, however, because the specific heat is negative for all energy values, no equilibrium states exist and the plasma never settles.  Because the system is forced, like a forced snow-pile (with continuous snow fall), as long as the metastable state persists, the energy input into the system from the externally applied currents balances the energy loss.  When the metastable state ceases, however, the rate of energy loss increases dramatically and the rate of energy input no longer balances the rate of energy loss.  This expansion causes a dramatic loss of core electron temperature which has been observed in experiment \cite{VonGoeler:1974}, a direct result of heat transport from the core to the outer edge via the runaway expansion mechanism.  Recovery occurs when the filaments expand far enough apart that the rate and magnitude of their collisions decreases, and the expansion slows.  Supposing that complete plasma disruption does not occur, the input current allows the energy to increase again, returning the plasma to a metastable state.  Hence, the instability is regular and repeated.  

This instability is a direct result of the confinement as Fig. \ref{fig:negspec2}, where confinement is weak, shows.  In the strong confinement regime, expansion decreases the potential because $v_1(r)$ increases from the core showing that density (\ref{eqn:density}) decreases; when confinement is weak, however, expansion increases it because $v_1(r)$ decreases from the core, and any further expansion increases the density at the edge of the confinement area.  In a real system, weakly confined plasmas do not persist for long before complete disruption however, so this regime is not useful for sustained fusion.

Negative specific heat has been found in neutral, two-component, electron-positron plasmas \cite{Kiessling:2003}.  This model, however, ignores Coulomb interactions between charged particles and does not take into account that the positively charged ion lines move far more slowly than the negatively charged electron lines.  For a neutral electron/ion plasma, a two-fluid magnetohydrodynamical model is appropriate.  At 100 million degrees K (the minimum target of tokamak reactors although we are not assuming a toroidal geometry here) electrons travel at a mean velocity of 40,000 kps while deuterons travel at ``only'' 600 kps.  Instabilities in tokamak plasmas such as the sawtooth instability may be related to the metastable state described in this paper \citep{Wesson:1999}.  Future research will focus on toroidal geometries to determine if this is likely.

\end{document}